\documentclass{emulateapj}
\usepackage{graphicx}

\newcommand{\degree}{$^{\circ}$}
\newcommand{\swift}{\emph{Swift}}

\newcommand{\cm}[1]{~cm$^{#1}$}

\newcommand{\e}[1]{10$^{#1}$}
\newcommand{\ee}[1]{$\times$10$^{#1}$}
\newcommand{\ergs}{~erg\,cm$^{-2}$\,s$^{-1}$}

\newcommand{\msun}{M$_{\odot}$}

\shorttitle{Short GRB jet-break}
\shortauthors{E. Troja et al.}

\begin{document}
\title{An achromatic break in the afterglow of the short GRB~140903A: \\
evidence for a narrow jet}

\author{E.~Troja\altaffilmark{1,2},
T. Sakamoto\altaffilmark{3},
S. B. Cenko\altaffilmark{2,4},
A. Lien\altaffilmark{5,2}, 
N. Gehrels\altaffilmark{2},
A.~J.~Castro-Tirado\altaffilmark{6,7},
R.~Ricci\altaffilmark{8},
J.~Capone\altaffilmark{1},
V.~Toy\altaffilmark{1},
A.~Kutyrev\altaffilmark{1,2},
N.~Kawai\altaffilmark{9},
A.~Cucchiara\altaffilmark{2,10}, 
A.~Fruchter\altaffilmark{10},
J.~Gorosabel\altaffilmark{1,11,12,$\dagger$},
S.~Jeong\altaffilmark{6,13},
A.~Levan\altaffilmark{14},
D.~Perley\altaffilmark{15},
R.~Sanchez-Ramirez \altaffilmark{6}, 
N.~Tanvir\altaffilmark{16}, 
S.~Veilleux \altaffilmark{1}
 }

\altaffiltext{1}{Department of Astronomy, University of Maryland, College Park, MD 20742, USA}	
\altaffiltext{2}{NASA, Goddard Space Flight Center, 8800 Greenbelt Rd, Greenbelt, Greenbelt, MD 20771, USA}	
\altaffiltext{3}{Department of Physics and Mathematics, College of Science and Engineering, Aoyama Gakuin University, 5-10-1 Fuchinobe, Chuo-ku, Sagamihara-shi, Kanagawa 252-5258, Japan}
\altaffiltext{4}{Joint Space-Science Institute, University of Maryland, College Park, MD 20742}
\altaffiltext{5}{Department of Physics, University of Maryland, Baltimore County, Baltimore, MD 21250, USA}
\altaffiltext{6}{Instituto de Astrof\'isica de Andaluc\'ia (IAA-CSIC), P.O. Box 03004, E-18008 Granada, Spain}
\altaffiltext{7}{Unidad Asociada Departamento de Ingenier\'ia y Sistemas Autom\'aticos, E.T.S. Ingenier\'ia Industrial, Universidad de M\'alaga, Campus de Teatinos, Arquitecto Francisco Penalosa, 6, 29010, M\'alaga, Spain}
\altaffiltext{8}{INAF-Istituto di Radioastronomia, Via Gobetti 101, I-40129 Bologna, Italy 0000-0003-4631-1528}
\altaffiltext{9}{Department of Physics, Tokyo Institute of Technology, 2-12-1 (H-29) Ookayama, Meguro-ku, Tokyo 152-8551, Japan}
\altaffiltext{10}{Space Telescope Science Institute, 3700 San Martin Drive, Baltimore, MD 21218}
\altaffiltext{11}{Unidad Asociada Grupo Ciencias Planetarias (UPV/EHU, IAA-CSIC), Departamento de F\'isica Aplicada I, E.T.S. Ingenier\'ia, Universidad del Pas Vasco (UPV/EHU), Alameda de Urquijo s/n, E-48013 Bilbao, Spain.}
\altaffiltext{12}{Ikerbasque, Basque Foundation for Science, Alameda de Urquijo 36-5, E-48008 Bilbao, Spain , Universidad del Pa\'is Vasco, Bilbao, Spain.}
\altaffiltext{13}{Sunkgkyunkwan University, 25-2 Sungkyunkwan-ro, Jongno-gu, 1398 Seoul, Korea.}
\altaffiltext{14}{Department of Physics, University of Warwick, Coventry,CV4 7AL, UK}
\altaffiltext{15}{Dark Cosmology Centre, Niels Bohr Institute, University of Copenhagen Juliane Maries Vej 30, 2100 Copenhagen, Denmark}
\altaffiltext{16}{Department of Physics and Astronomy, University of Leicester, Leicester, LE1 7RH, UK}		 
\altaffiltext{$\dagger$}{Deceased}.

\begin{abstract}

We report the results of our observing campaign on GRB~140903A, 
a nearby ($z=0.351$) short duration ($T_{90}$$\sim$0.3~s) gamma-ray burst 
discovered by {\it Swift}.
We monitored the X-ray afterglow with {\it Chandra} up to 21 days after the burst, 
and detected a steeper decay of the X-ray flux after $t_j$$\approx$1~day. 
Continued monitoring at optical and radio wavelengths showed a similar decay in flux
at nearly the same time, and we interpret it as evidence of a narrowly collimated jet. 
By using the standard fireball model to describe
the afterglow evolution, we derive a jet opening angle $\theta_j$$\approx$5~deg and a collimation-corrected total energy release $E$$\approx$2\ee{50} erg.
We further discuss the nature of the GRB progenitor system. 
Three main lines disfavor a massive star progenitor: the properties of the prompt gamma-ray emission, 
the age and low star-formation rate of the host galaxy, and the lack of a bright supernova. 
We conclude that this event was likely originated by a compact binary merger. 
\end{abstract}

\keywords{X-rays: bursts; gamma ray burst:  individual (GRB~140903A);}

\section{Introduction}\label{sec:intro}

Gamma-ray bursts (GRBs) are produced by a highly relativistic outflow collimated into jets. The angular size of the outflow is therefore a key ingredient in determining the true energy release and the event rate. These parameters provide a crucial test for any progenitor and central engine model.

Measuring the collimation of short duration GRBs, i. e. those lasting less than 2~s \citep{kouvel93},  is not only a primary interest of the GRB field, 
but has a broader impact. 
Growing observational evidence connects short GRBs with compact  binary mergers \citep[][and references therein]{gehrels05,tanvir13,berger14,yang15}, 
which are among the most promising sources of gravitational wave (GW) radiation \citep{thorne87, ligo16}.  Therefore, the degree of collimation of 
short GRBs is a critical input for inferring the true rate of binary mergers, the expected detection rate of advanced LIGO and Virgo \citep{abadie10}, and for estimating our chances to observe the electromagnetic counterpart of a GW source (\citealt{ligoem16,troja16}).

Observationally, the beamed geometry leaves a clear signature
in the afterglow temporal evolution, manifesting itself as an {\em achromatic} light curve break
(known as ``jet-break''), visible on timescales of 
$\sim$days-weeks after the explosion \citep{rhoads99}.
At early times (hours after the explosion), the evolution of
the afterglow is the same as for a spherical explosion.
However, later on, the jet edges
become visible causing the observed flux to rapidly fall off \citep{vaneerten10,vaneerten13}. 
For a jet expanding into a homogeneous ambient medium such steepening takes place at a time $t_j$\,$\propto$\,$\theta_j^{8/3}$ \citep{sari99,vaneerten10},
when the outflow is decelerated down to a bulk Lorentz factor $\approx$\,$\theta_j^{-1}$, 
where $\theta_j$ is the jet half-opening angle.
The detection of a jet-break in the afterglow light curve 
is therefore an important diagnostic tool for constraining
the outflow geometry, and the burst energetics.

In the case of short bursts, the faintness of their afterglows often hampers the search for jet-breaks.  
Only a small fraction of short GRBs have been detected at optical or radio wavelengths, and often sampled too poorly to 
meaningfully constrain the afterglow temporal evolution \citep{kann11,davanzo14}.  
\citet{nicuesa12} presented good evidence for an achromatic steepening in the optical/NIR light curve of the short GRB~090426. 
However, the classification of this burst is rather ambiguous \citep{antonelli09,levesque10}, and it was proposed that the event was more likely an interloper, originated by a massive star progenitor \citep{thone11, virgili11, nicuesa12}

Candidate jet-breaks have been identified in several X-ray afterglows of short GRBs (\citealt{burrows06}, Soderberg et al.~2006, \citealt{stratta07, fong12, coward12,zhang15}), 
however their interpretation as jet-breaks remain quite controversial.  
Several studies suggest that the X-ray light curves may be shaped by a persistent energy injection from the central engine \citep{fan06,ctn11,rowlinson13} rather than by external shock emission \citep{mesre97}. 
In this scenario the sharp decay of the X-ray flux is attributed to the rapid turn-off of the energy source rather than to the outflow geometry, and no collimation is needed to explain the observed light curves.  
Indeed,  in the small sample of events with simultaneous optical and/or radio coverage (e.g. GRB090510, \citealt{max10}; GRB130603B, \citealt{tanvir13})
the observed temporal breaks appear to be chromatic rather than frequency-independent.  The jet-break interpretation, to still hold, would require an alteration of the basic jet model, such as a two-component jet \citep{corsi10}, evolving shock parameters \citep{max10}, or the presence of additional emission components \citep[e.g.][]{gao15}.
  
In this paper, we present our multi-wavelength campaign of the short GRB~140903A
which revealed an achromatic break in its afterglow light curve.
Through the analysis of the broadband data, we show that the observed emission is fully 
consistent with the standard forward shock model, and requires a narrowly collimated outflow. 
A previous analysis of this event, based on {\it Swift} observations,  
did not detect the presence of a jet-break in the X-ray data \citep{fong15}.
Our addition of deep, late-time {\it Chandra} observations is indeed critical 
for the jet-break detection and its characterization. 
We further investigate the GRB classification and the nature of its progenitor, and conclude that
this event is a {\it bona fide} short GRB, likely originated by a compact binary merger. 
The paper is organized as follows: our observations and data reduction procedures
are detailed in $\S$~\ref{sec:obs}; we present our analysis of the 
GRB prompt emission, its afterglow and host galaxy in $\S$~\ref{sec:data};
our results are discussed in $\S$~\ref{sec:disc}.
Throughout the paper, times are referred to the {\it Swift} trigger time,
and the phenomenology of the burst is presented in the observer's frame.
We employ a standard $\Lambda$CDM cosmology with 
$H_0 = 67.8$\,km\,s$^{-1}$\,Mpc$^{-1}$, 
$\Omega_M$ = 0.308, and $\Omega_{\Lambda}$=0.692 \citep{planck15}.
Unless otherwise stated, errors are given at the 68\% confidence level
for one interesting parameter, and upper limits are reported at the 3\,$\sigma$ confidence level.

\newpage

\section{Observations and Data Reduction}\label{sec:obs}
\subsection{Swift BAT and XRT}\label{swift}
GRB~140903A triggered the \swift~Burst Alert Telescope 
\citep[BAT;][]{bat05} at 15:00:30 UT on 3rd September, 2014 \citep{gcn14}. 
The {\it Swift} X-ray Telescope \citep[XRT;][]{xrt05} began settled observations
of the GRB field 74~s after the BAT trigger, and monitored the X-ray afterglow
during the following 3 days, until the source faded below the detector 
sensitivity threshold. 
The XRT data comprise 24~ks acquired in Photon Counting (PC) mode.

BAT and XRT data were processed using the {\it Swift} software package
distributed within HEASOFT (v.~6.17). We used the latest release of the BAT and XRT
Calibration Database, and followed standard data reduction procedures.

\subsection{Chandra}\label{cxo}
The {\it Chandra} X-Ray Observatory performed two Target of Opportunity (ToO)
observations in order 1) to precisely localize the X-ray afterglow (PI: T.~Sakamoto),
and 2) to characterize its late-time temporal evolution,
and search for a possible jet-break (PI: E.~Troja). 
Our first observation 
\dataset[ADS/Sa.CXO#obs/15873]{(ObsId 15873)}
started 3\,d after the burst,
and observed the field for a total exposure of 19.8\,ks.
Our second {\it Chandra} observation 
\dataset[ADS/Sa.CXO#obs/15986]{(ObsId 15986)}
was performed on 2014, Sep 18.
for a total exposure of 59.3\,ks.
{\it Chandra} data were reduced using version 
4.6.1 of the CIAO software with CALDB version 4.6.3.
Events from the GRB afterglow were selected using a source extraction radius of 
2~pixels, and the derived count rates were corrected for vignetting effects and
Point Spread Function (PSF) losses. 
The background contribution was estimated from an annular, 
source-free region centered on the afterglow position.

The GRB afterglow is detected at both epochs.
In our first {\it Chandra} observation we detect 80 net source counts
in the 0.5-8.0 keV energy band. 
We corrected the native {\it Chandra} astrometry by aligning our X-ray and optical  images (see $\S$~\ref{dct}).
Based on the match of five bright X-ray and optical sources, 
we determine a refined X-ray (J2000.0) position of  
$\alpha = 15^{\mathrm{h}} 52^{\mathrm{m}} 03.273^{\mathrm{s}}$, 
$\delta = +27^{\circ} 36\arcmin 10\farcs83$ with an error radius of 0.4\arcsec (90\% confidence level).
In our second and last {\it Chandra} observation only 6 counts are measured
at the source position, corresponding to a detection significance $>$99.99\% \citep{kraft91}.


\begin{figure*}[!t]
\centering
\vspace{0.2cm}
\includegraphics[angle=0,scale=0.46]{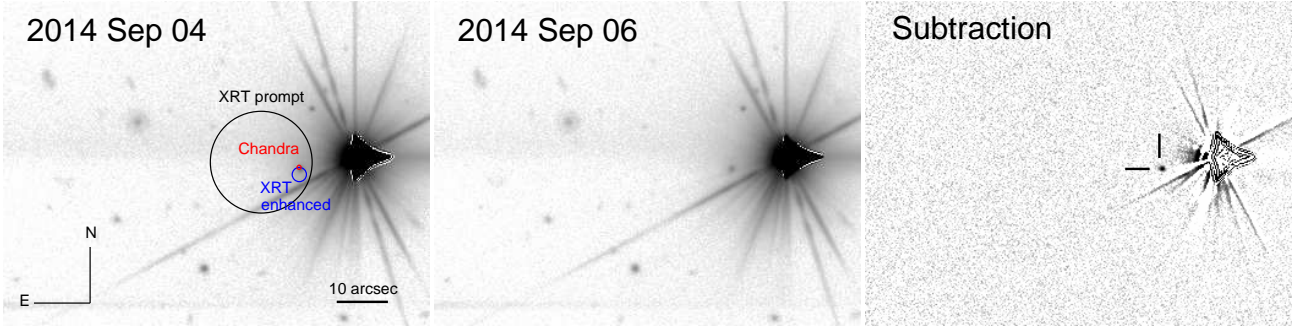}
\caption{DCT $r$-band observations of the field of GRB~140903A, taken at 0.5 days (\textit{left panel})
and 2.5 days (\textit{middle panel}) after the burst. 
The black circle shows the initial XRT afterglow localization.
The blue and red circles show the XRT enhanced and the refined {\it Chandra} positions, respectively.
\textit{Right panel}: image subtraction of the two previous panels,  showing the residual afterglow light.
}
\vspace{0.2cm}
\label{fig:finder}
\end{figure*}


 \subsection{Discovery Channel Telescope}\label{dct}
We initiated an observing campaign with the Large Monolithic Imager
(LMI) mounted on the 4.3\,m Discovery Channel Telescope (DCT) in Happy Jack,
AZ. Observations in the $griz$ filters started on 2014 Sep 04 at 3.17 UT, 
approximately 12 hours after the {\it Swift} trigger, 
and continued to monitor the field for the next 3 weeks.
Late-time images in the $r$ and $i$ filters were acquired on 2016 March 17 (561 days after the burst)
and used as templates for image subtraction. 
Standard CCD reduction techniques (e.g., bias subtraction, flat-fielding, etc.) were
applied using a custom IRAF\footnote{IRAF is distributed by the National Optical 
Astronomy Observatory, which is operated by the Association of Universities for 
Research in Astronomy (AURA) under cooperative agreement with the National Science 
Foundation.} pipeline.  Individual short (10--20\,s) exposures were aligned with
respect to astrometry from the Sloan Digital Sky Survey (SDSS; \citealt{aaa+14})
using SCAMP \citep{b06} and stacked with SWarp \citep{bmr+02}.

As shown in Figure~\ref{fig:finder}, the field of GRB\,140903A is quite complex: 
the optical afterglow lies on top of a relatively bright host galaxy (see below),
and only 12\arcsec~away from an extremely bright ($V \approx 9$\,mag) star.
In order to extract the afterglow brightness from our DCT images, 
we performed digital image subtraction 
with the High Order Transform of PSF ANd Template
Subtraction (HOTPANTS; \citealt{hotpants}).
The resulting photometry, calibrated with respect to nearby point sources
from SDSS, is presented in Table~\ref{tab:opt}.  The transient is detected with high
significance in our first epoch at $\Delta t = 12.5$\,hr, and possibly at 2.5~days,
although the significance of this last detection is only marginal ($\lesssim$3\,$\sigma$). 

Using the images from 2016 March 17, we measure the following magnitudes for the
underlying host galaxy: 
$r^{\prime} = 20.58 \pm 0.09$,
and $i^{\prime} = 20.12 \pm 0.05$.
From earlier observations we also measure $g^{\prime} = 21.97 \pm 0.16$, and $z^{\prime} = 19.66 \pm 0.08$, 
although we caution that these fluxes may include some afterglow contribution.
The host is unresolved in all of our DCT images (seeing ranging from 0\farcs8--2\farcs0).  
Using astrometry from nearby SDSS point sources for reference, we measure a 
(J2000.0) position of $\alpha = 15^{\mathrm{h}} 52^{\mathrm{m}} 
03.278^{\mathrm{s}}$, $\delta = +27^{\circ} 36\arcmin 10\farcs68$.  The excess
afterglow flux measured in our subtracted images is consistent with this location,
within the estimated uncertainty of our astrometric tie ($\approx$100\,mas in 
each coordinate).

\begin{table*}[!tbh]
\caption{Log of Optical and Near-IR Observations} 
\begin{center}       
\begin{tabular}{cccccccc} 
\hline
\hline
Date & Time Since Burst & Telescope & Instrument &  Filter & Exposure Time & Afterglow Magnitude$^{\rm a}$ & Host Magnitude$^{\rm a}$ \\
 (UT) & (d) &  & & & (s) & (AB) & (AB)  \\
\hline
2014 Sep 4.13 & 0.51 &  DCT & LMI & $r^{\prime}$ & 300 & $21.56 \pm 0.08$ & $\cdots$ \\
2014 Sep 4.15 & 0.53 & DCT & LMI & $r^{\prime}$ & 300 & $21.63 \pm 0.06$ &$\cdots$ \\
2014 Sep 4.22 & 0.60 & Gemini & GMOS & $i^{\prime}$ & 120 & $21.33 \pm 0.05$ &$\cdots$ \\
2014 Sep 5.18 & 1.55 & DCT & LMI & $r^{\prime}$ & 630 & $> 21.2$ &  $\cdots$ \\
2014 Sep 5.26 & 1.63 & Gemini & GMOS & $i^{\prime}$ & 600 &  $22.99 \pm 0.13$  & $\cdots$  \\
2014 Sep 6.12 & 2.50 & DCT & LMI & $r^{\prime}$ & 600 & $>$22.3$^{\rm b}$ &  $\cdots$ \\
2014 Sep 6.15 & 2.52 & DCT & LMI & $i^{\prime}$ & 600 & $> 22.0 $ &  $\cdots$ \\
2014 Sep 6.17 & 2.55 & DCT & LMI & $g^{\prime}$ & 600 & $\cdots$ &  21.97 $\pm$0.16 \\
2014 Sep 6.20 & 2.57 & DCT & LMI & $z^{\prime}$ & 600 & $\cdots$ &  19.66 $\pm$0.08 \\
2014 Sep 6.84 & 3.22 & LT & IO:O & $z^{\prime}$ & 900 &  $>22.5$ &  $\cdots$ \\
2014 Sep 8.12 & 4.49 & DCT & LMI & $r^{\prime}$ & 600 & $>$22.9 &  $\cdots$  \\
2014 Sep 8.14 & 4.52 & DCT & LMI & $i^{\prime}$ & 340 & $>21.4$ &  $\cdots$ \\
2014 Sep  8.84 & 5.22 & LT & IO:O & $z^{\prime}$ & 1200 & $\cdots$ &  19.64$\pm$0.13 \\
2014 Sep 13.82 & 10.20 & CAHA & $\Omega_{2000}$ & $J$ & 720 & $\cdots$ & 18.92$\pm$0.05 \\
2014 Sep 13.84 & 10.22 & CAHA & $\Omega_{2000}$ & $H$ & 1200 & $\cdots$ & 18.57$\pm$0.07 \\
2014 Sep 13.85 & 10.23 & CAHA & $\Omega_{2000}$ & $K_s$ & 1800 & $\cdots$ & 18.25$\pm$0.05 \\
2014 Sep 23.11 & 19.49 & DCT & LMI & $i^{\prime}$ & 600 & $> 22.9 $ &  $\cdots$ \\
2016 Mar 17.92 & 561   & DCT & LMI  & $r^{\prime}$ & 880 & $\cdots$ &  20.58$\pm$0.09 \\
2016 Mar 18.19 & 561   & DCT & LMI  & $i^{\prime}$ & 880 & $\cdots$ &  20.12$\pm$0.05 \\
2016 Apr  02.48 & 577   & Gemini & GMOS & $i^{\prime}$ & 600 & $\cdots$ &  20.28$\pm$0.09 \\
\hline
\end{tabular}

\footnotetext{~Values not corrected for Galactic extinction.}
\footnotetext{~A faint excess ($r$=23.11$\pm$0.36) is visible in the residual difference image. Its significance is only marginal ($\lesssim$3\,$\sigma$), and
we cannot exclude that it is an artifact of the subtraction method.}
\end{center}
\label{tab:opt}
\end{table*}

\subsection{Liverpool and Calar Alto Telescopes}\label{spain}

Near-IR images were acquired in $zJHK_s$-bands using the 2.0m Liverpool (LT) and the 3.5m Calar Alto telescopes (CAHA). 
The LT images were taken in the $z$-band with the IO:O camera, which provides a $10.0^{\prime}\times10.0^{\prime}$ field of view and a
$0.3^{\prime\prime}$ pixel scale. The CAHA data were acquired in the $JHK_s$-bands with the $\Omega_{2000}$ instrument, yielding a
$15.4^{\prime}\times15.4^{\prime}$ field of view and a $0.45^{\prime\prime}$ pixel scale. 
In order to reduce the contamination of the nearby bright star,  
these observations were taken in relatively short (20 s - 30 s) exposures.
The reduction followed standard steps;
bad pixel masking, bias and flat field correction, sky subtraction, plus
stacking, performed by calling on IRAF tasks \citep{Tody93}.
The resulting photometry, calibrated with respect to nearby point sources
from SDSS and 2MASS \citep{scs+06}, is presented in Table~\ref{tab:opt}.
We used the offsets from \citet{br07} to convert the 2MASS Vega magnitudes
to the AB system. 

 \subsection{Gemini Imaging}\label{gemini}
 
We imaged the field of GRB\,140903A with the Gemini Multi-Object Spectrograph (GMOS; \citealt{hja+04} on the 8\,m Gemini North telescope.
We obtained a single 120\,s $i^{\prime}$ image beginning at  05:24 UT on 2014 Sep 4 ($\Delta t = 14.4$\,hr), and a dithered sequence
of 10 $\times$ 60\,s $i^{\prime}$ exposures at a mean epoch of $\Delta t = 39.2$\,hr on 2014 Sep 5.  
Our last observation was performed on 2016 April 2 and used as template for image subtraction.
The images were reduced in the standard manner using the \texttt{gemini} IRAF package.
We performed digital image subtraction on the GMOS images using the same 
analysis methods as was used for the DCT images (Sect.~\ref{dct}).  In the subtracted frame transient emission
is clearly detected at an offset of 96$\pm$44~mas from the galaxy's center. 
At a redshift $z$=0.351 this corresponds to a physical projected offset of 0.5$\pm$0.2~kpc.  
For the transient component we infer $i^{\prime} = 21.33 \pm 0.05$\,mag in our first epoch,
and $i^{\prime} = 22.99 \pm 0.13$\,mag in the second epoch.  This implies a steep temporal
decay with slope $\alpha_o$=1.54$\pm$0.15 between the two observations. 



\begin{figure}[!b]
\centering
\vspace{0.6cm}
\includegraphics[angle=0,scale=0.34]{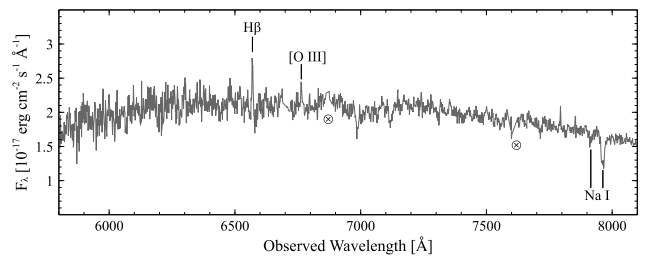}
\caption{Gemini GMOS spectrum of GRB~140903A and its host galaxy, acquired 14.6~hrs after the burst.
The positions of detected emission and absorption lines are indicated.
Crossed circles mark the position of strong telluric features.
}
\label{fig:gmos}
\end{figure}


 \subsection{Gemini Spectroscopy}\label{geminispec}
We obtained a series of spectra of the afterglow+galaxy with 
GMOS beginning at 05:34 UT on 2014 Sep 4 ($\Delta t = 14.6$\,hr).  GMOS
was configured with the R400 grating and a central wavelength of 600\,nm,
providing coverage from $\lambda \approx 4000$--8000\,\AA\ with a 
resolution of $\approx1000$.  
We restricted our analysis to $\lambda$$>$5500\,\AA\  due to the poor signal-to-noise
of the spectrum at lower wavelengths.

The resulting spectrum is plotted in Figure~\ref{fig:gmos}.  The strongest
(non-telluric) feature is a broad (FWHM$\approx$15~\AA) absorption line at $\lambda \approx
7963$\,\AA, along with a weaker (but still broad, FWHM$\approx$10~\AA) absorption line
at $\lambda \approx 7915$\,\AA.  We interpret these features as
corresponding to \ion{Na}{1} with $z \approx 0.35$.
We also detect narrow emission lines at $\lambda = 6569.6 \pm 0.5$\,\AA\
and $\lambda = 6763.7 \pm 0.6$\,\AA, which correspond to H$\beta$ and
[\ion{O}{3}] at $z = 0.351 \pm 0.001$, which we adopt for the redshift
of the host\footnote{The weaker [\ion{O}{3}] $\lambda$4959 line falls
in the chip gap at $z$ = 0.351.}.  Weak absorption features corresponding
to \ion{Ca}{2} H+K are also visible at this redshift, though with 
marginal significance.

 \subsection{GTC Spectroscopy}\label{gtc}
 
Further optical spectroscopy of the host galaxy was performed using OSIRIS (Optical System for Imaging and low Resolution Integrated Spectroscopy; \citealt{cepa2000}) at the 10.4 m GTC. Observations started on Oct 03, 2014, i.e. $\sim$ 30.2 d after the trigger, using the R1000B grism (2$\times$600 s exposures) and R2500I VPH (3$\times$600 s exposures).  The spectra covered the 3600--7800\,\AA\ range at a resolution of $\approx$1000 and the 7300--10,000\,\AA\ range at resolution $\approx$2500.
The 1.0" slit was positioned on the location of the host galaxy and 2 $\times$ 2 binning mode was used for data acquisition. The obtained spectra were reduced and calibrated following standard procedures using custom tools based in IRAF and Python. Spectra were flux calibrated using the spectrophotometic standard star GD248, which was observed during the same night with a 2.52\arcsec~slit. In order to account for slit losses, we renormalized the flux of the source to match the DCT magnitudes shown in Table~\ref{tab:opt}. Acquisition images were not usable due to the nearby saturated star.

Although close to a skyline, H$\alpha$ is clearly detected in the red spectrum at $\lambda = 8862.1 \pm 0.8$\, \AA,
consistent with the redshift from GMOS (Section~\ref{geminispec}).
No emission lines are visible in the blue grism spectrum. 
This may be due to the presence of dust, also suggested by a clear spectral curvature towards the short wavelengths.

\begin{table}[!t]
\caption{Log of Radio Observations} 
\begin{center}       
\begin{tabular}{cccc} 
\hline
\hline
Date & Time Since Burst & Frequency &  Flux  \\
 (UT) & (d) &  ( GHz)&  ($\mu$Jy)  \\
\hline
2014 Sep 04.06 & 0.44 & 6.1 & 118$\pm$11  \\
2014 Sep 06.13 & 2.51 &  6.1 & 203$\pm$13 \\
                          &         &  9.8 & 153$\pm$10 \\
2014 Sep 07.92 & 4.30 & 6.1 & 141$\pm$17\\
2014 Sep 12.89 &  9.27 & 6.1 & 90$\pm$20  \\
                          &          & 9.8 & $<$75 \\
2014 Sep 21.88 &  18.26 & 6.1 & $<$130 \\
\hline
\end{tabular}
\end{center}
\label{tab:jvla}
\end{table} 

 \subsection{Jansky Very Large Array}\label{vla}

GRB~140903A was observed with the Jansky Very Large Array (VLA) at both 6.1~GHz (C-band) and at 9.8~GHz (X-band).
Observations started $\sim$10~hrs after the burst, and periodically monitored the source for 18 days \citep{fong15}. 
Radio data were downloaded from the public NRAO archive, and reduced using the Common Astronomy Software Applications (CASA) v.~4.5.2 package. 
After standard calibration and basic flagging, we visually inspected the data and applied further screening when needed. Galaxies 3C286 and J1609+2641 were used as flux and phase calibrators, respectively.   The log of radio observations is reported in Table~\ref{tab:jvla}. Our values are slightly higher, but largely consistent with those reported by \citet{fong15}.  A simple power-law fit to the data yields decay slopes $\alpha_{6GHz}$=0.63$^{+0.14}_{-0.12}$ and $\alpha_{9.8GHz}$$>$0.5 for $t$$>$1 d. This does not take into account the possible effects of interstellar scintillations (ISS), which we model in Section~\ref{sec:agmod}.

\section{Data Analysis}\label{sec:data}
\subsection{Gamma-ray data}\label{sec:bat}

\begin{figure}[!b]
\centering
\vspace{0.2cm}
\includegraphics[angle=0,scale=0.33]{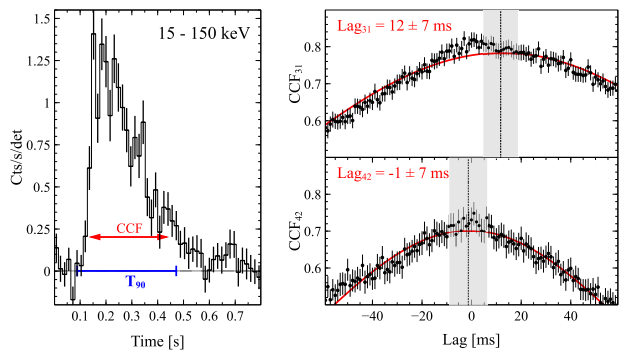}
\caption{{\it Left panel}: BAT light curve of GRB~140903A
in the 15--150~keV energy band.
The T$_{90}$ time interval, and the time interval used for the 
cross-correlation function (CCF) analysis are shown.
{\it Right panel}: CCFs between the standard BAT energy bands.
The best fit gaussian function is reported as a solid line.
The lag value and its uncertainties are indicated
by the vertical shadowed region.
}
\vspace{0.2cm}
\label{fig:ccf}
\end{figure}

The prompt emission consists of a main Fast Rise Exponential Decay (FRED) pulse,
with a duration of $T_{90}$=\,0.30$\pm$0.03\,s in the 15-350 keV band (Figure~\ref{fig:ccf}, left panel).
The time-averaged spectrum, from T+0.09 to T+0.47, shows that the prompt
emission is well described ($\chi^2$=44 for 57 degrees of freedom) 
by a simple power-law with $\Gamma$=1.99$\pm$0.08.
According to this best fit model, the burst fluence in the observed 15-150~keV energy band
is (1.35$^{+0.07}_{-0.05}$)\ee{-7}\,erg\,cm$^{-2}$, which,
at a redshift $z$=0.351, corresponds to an isotropic-equivalent energy 
of $E_{\gamma,\rm iso}$=(6.0$\pm$0.3)\ee{49}\,erg.
Due to the narrow BAT energy bandpass, this only places a lower limit to the bolometric 
energy release. 
However, for a typical GRB spectrum \citep{band93}, the measured soft photon index indicates 
that the spectral peak lies close to or within the BAT energy range \citep{taka09}.
In this case, the bulk of the emission mainly falls within the observed range, and the
derived value of $E_{\gamma,\rm iso}$ represents a good estimate of the total energy 
radiated in the prompt emission.


Spectral lags were calculated by cross-correlating the light curves in the standard BAT channels: 
1 (15-25 keV), 2 (25-50 keV), 3 (50-100 keV), 4 (100-350 keV). 
We followed the method outlined by \citet{ukwatta12} and, in order to increase the 
signal-to-noise in the higher energy channels, performed the analysis on 
non mask-weighted lightcurves, each with a 4 ms time resolution. 
The derived lags are $\tau_{31}$=12$^{+7}_{-7}$\,ms and $\tau_{42}$=\mbox{-1$^{+7}_{-7}$\,ms}, where the quoted uncertainties 
were derived  by Monte Carlo simulations. The results of our lag analysis are shown in Fig.~\ref{fig:ccf} (right panel).

We also searched for temporally extended emission following the main burst, 
but no significant signal was found. By assuming a power-law spectrum with photon
index $\Gamma$=2, we set a 3\,$\sigma$ upper limit of 8\ee{-10}\,\ergs~(15-50 keV) in the
time interval 10-100 s. This is consistent with the MAXI  upper limit of
8.4\ee{-10}\,\ergs~in the 4-10~keV energy band \citep{maxigcn}. 

\begin{figure*}[!t]
\centering
\vspace{0.5cm}
\includegraphics[angle=0,scale=0.65]{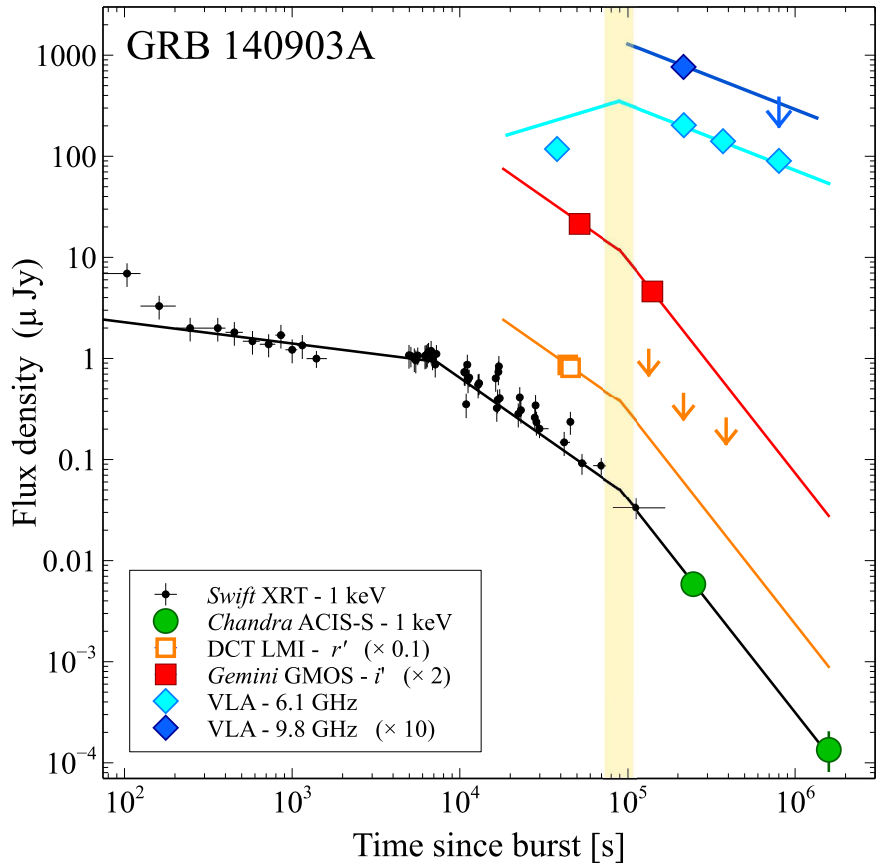}
\caption{Afterglow light curves of GRB~140903A,
combining X-ray data from the {\it Swift}/XRT (small circles), and 
the {\it Chandra}/ACIS-S (large circles), optical data from DCT (open squares), Gemini (filled squares), 
and radio data from the VLA (diamonds). 
Error bars are~1~$\sigma$, arrows denote 3~$\sigma$ upper limits. The best fit temporal model is shown as a solid line.
The vertical band marks the time of the jet-break. }
\vspace{0.1cm}
\label{fig:xlc}
\end{figure*}


\begin{figure*}[!t]
\centering
\vspace{0.5cm}
\includegraphics[angle=0,scale=0.7]{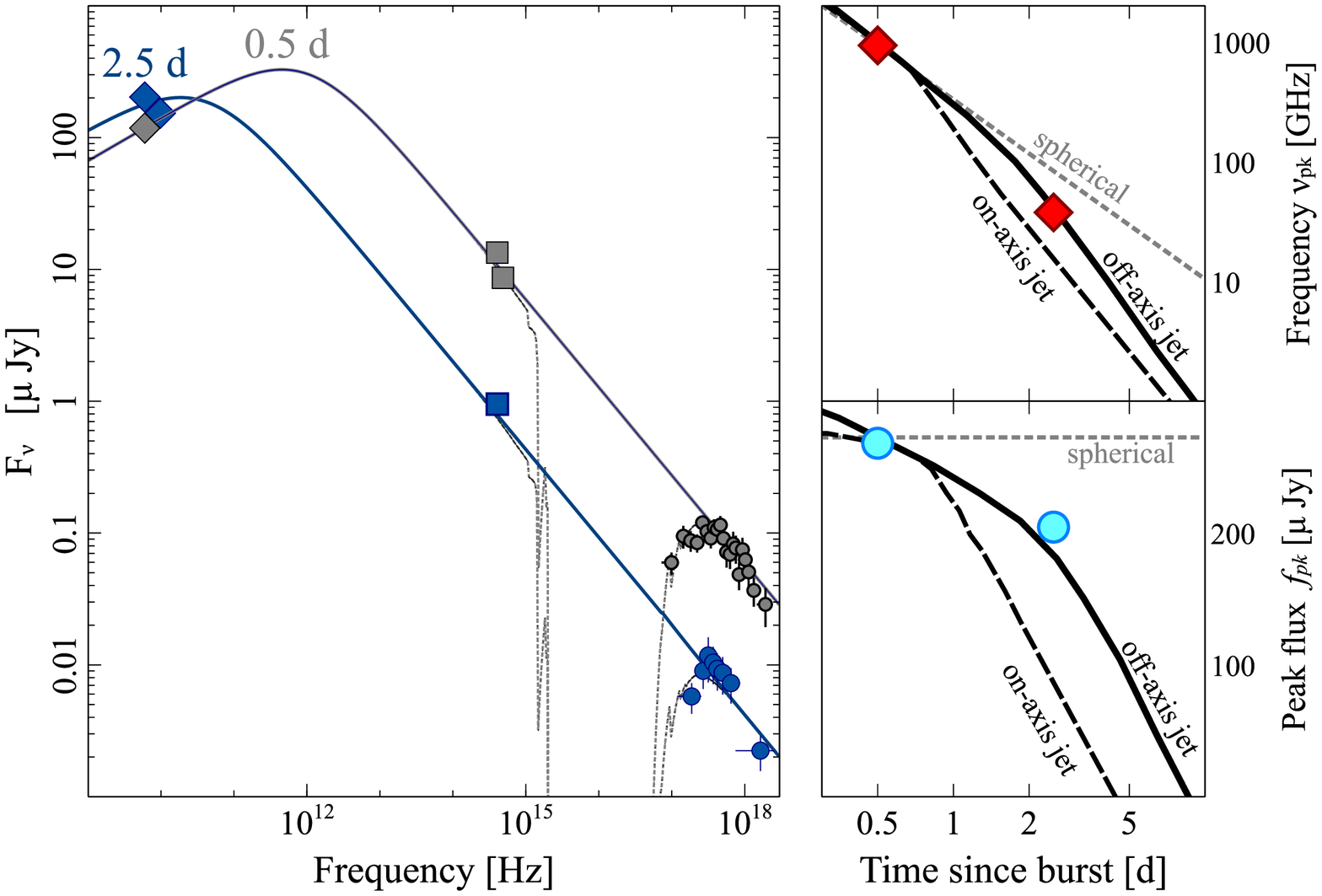}
\caption{{\it Left panel:} Afterglow spectral energy distribution at two different epochs, $t_1$=0.5~d before the jet-break, and $t_2$=2.5~d after the jet-break.
We fit the broadband spectrum with a smoothly broken power-law, our best fit models are shown by the solid lines. The thin dotted lines show the effects of absorption and extinction. {\it Top right panel:} Temporal evolution of the peak frequency across the jet-break. 
We report the expected behavior for three different models: the spherical fireball (dotted line),  a narrow jet ($\theta_{jet}$=0.1~rad)  seen on-axis (dashed line), and seen slightly off-axis ($\theta_{obs}/\theta_{jet}$=0.6; solid line). Our measurements, indicated by the red diamonds, agree well with the off-axis jet model. 
 {\it Bottom right panel:} Same as above but for the spectral peak flux. Also in this case, our  derived values (cyan circles) agree well with the trend expected from an off-axis jet.
}
\vspace{0.1cm}
\label{fig:sed}
\end{figure*}


\subsection{X-ray data}\label{sec:xray}

\subsubsection{Spectral analysis}\label{sec:pha}

The afterglow spectral parameters were derived from
the time-averaged XRT/PC spectrum (from 100~s to 110~ks).
We binned the data in order to have at least 1 count
per spectral channel, and performed the fit within XSPEC \citep{xspec} ~v.12.9.0 by minimizing
the Cash statistic.
The spectrum is well described by an absorbed power law model
(W-stat=329 for 359 degrees of freedom, d.o.f.). 
The best fit parameters are a photon index $\Gamma_X$=1.66$\pm$0.09,
and an absorbing column 
$N_{\rm H, int}$($z$=0.351)=(1.3$\pm$0.4)\ee{21}\,cm$^{-2}$, 
in excess to the Galactic value $N_{\rm H,Gal}$=2.9\ee{20}\,cm$^{-2}$
in the burst direction \citep{kalberla05}. 
The adopted value is consistent within errors with the $N_{\rm H,Gal}$=$3.3\times10^{20}$\,cm$^{-2}$ estimated by \citet{willi13}.

In our first {\it Chandra} observation, the source spectrum is well fit (W-stat=55 for 58 dof) by 
an absorbed power law model with $\Gamma_X$=1.8$\pm$0.2,
 and $N_{\rm H, int}$=1.3\ee{21}\,cm$^{-2}$, fixed at the value of the XRT best fit. 
In our second and last {\it Chandra} observation the low number of counts
prevents any spectral analysis. 
As the hardness ratios of the two {\it Chandra} observations are consistent within
the uncertainties, the same spectral parameters were adopted to estimate the observed flux.

For the best fit parameters quoted above,
we derived an unabsorbed energy conversion factor (ECF)
of (4.8$\pm$0.2)\ee{-11}\,erg\,cm$^{-2}$\,cts$^{-1}$
for the {\it Swift}/XRT data, and of 
(1.40$\pm$0.15)\ee{-11}\,erg\,cm$^{-2}$\,cts$^{-1}$
for the {\it Chandra} ACIS-S data.

\subsubsection{Temporal Analysis}\label{sec:lc}

The X-ray light curve was binned to have a minimum of 15 counts
in each temporal bin. The observed count-rates were converted 
into flux units by using the ECFs derived in Section~\ref{sec:pha}, 
and by propagating the relative uncertainties.
We modeled the afterglow temporal decay with a series of
power-law segments ($f_X \propto t^{-\alpha_i}$) and minimized
the $\chi^2$ statistics to obtain the best fit to the data.
The afterglow displays a shallow decay phase with temporal
index $\alpha_1$$\sim$0.2, which steepens
to $\alpha_2$$\sim$1.1 after $t_{bk,1}$$\sim$7~ks.
Our first {\it Chandra}/\mbox{ACIS-S} data point lies below the predictions based on the {\it Swift}/XRT dataset, 
hinting at a second temporal break in the light curve.
However, the combined XRT/ACIS-S dataset could be reasonably well described
 by adopting a steeper temporal index $\alpha_2$$\sim$1.5 for the final
power-law decay, and no additional break.
A second {\it Chandra} observation was therefore executed in order to 
distinguish between the two models.
This last measurement confirms the presence
of an additional break in the X-ray light curve at a time $t_j$$\approx$1~d,
and allows us to constrain the slope of the final decay to $\alpha_3$$\sim$2.1
The best fit temporal models are summarized in Table~\ref{tab:bfit}.
The X-ray light curve and our best fit model are presented in Fig.~\ref{fig:xlc}, 
and compared to the optical (Table~\ref{tab:opt}) and radio measurements  (Table~\ref{tab:jvla}) in order to highlight the achromatic nature
of the last temporal break $t_j$, which we interpret as the jet-break time. 

\begin{table}[!bht]
\caption{Afterglow light curve fit parameters} 
\begin{center}  
\resizebox{\columnwidth}{!}{
\begin{tabular}{ccccccc}
\hline
\hline
Band &   $\alpha_1$ & $t_{bk,1}$ & $\alpha_2$ & $t_{bk,2}$ & $\alpha_3$ &  $\chi^2$ / dof \\
          &                     & (ks) &  & (ks) &  &   \\
\hline
X & 0.20$\pm$0.02 & 7.3$^{+0.6}_{-0.9}$ &  1.06$^{+0.07}_{-0.11}$ & 69$^{+17}_{-12}$ & 2.11$^{+0.22}_{-0.07}$ & 43 / 46  \\
O & 1.54$\pm$0.15 & -- &  -- & -- & -- & --  \\
X+O & 0.21$\pm$0.02 & 7.9$^{+1.0}_{-0.9}$ &  1.16$^{+0.10}_{-0.03}$ & 89$^{+11}_{-12}$ & 2.1$^{+0.2}_{-0.2}$ & 49 / 48  \\
R (6.1~GHz)  & -0.5$^a$ & 89$^{+11}_{-12}$ &  0.63$\pm$0.14 & -- & -- & --  \\
R (9.8~GHz) & $>$0.5 & -- &  -- & -- & -- & --  \\
\hline
\end{tabular}
}
\end{center}
\label{tab:bfit}
\footnotetext{The temporal slope was held fixed at the value predicted by the standard fireball model for $\nu_{sa}$$<$$\nu$$<$$\nu_m$.}
\end{table}

\subsection{Afterglow Spectral Energy Distribution}

In order to study the spectral evolution across the temporal break $t_{j}$ detected in X-rays,
we extracted the afterglow spectral energy distribution (SED) at two different epochs,
$t_1$=0.5~d ($< t_{j}$) and  $t_2$=2.5~d ($> t_{j}$). 
These times were selected in order to maximize the simultaneous coverage
at different wavelengths. 

 Optical fluxes were derived by the best fit temporal model in Table~\ref{tab:bfit}, and
corrected for Galactic extinction in the GRB direction ($E_{B-V}$$\approx$0.03; \citealt{ebv98}).
A power-law fit ($f_{\nu}$\,$\propto$\,$\nu^{-\beta}$) to the optical 
and X-ray data yields spectral slopes  $\beta_{OX}$=0.72$\pm$0.05 at $t$=$t_1$,
$\beta_{OX}$=0.76$\pm$0.12 at $t$=$t_2$,
significant intrinsic absorption $N_H$=(1.8$\pm$0.4)\ee{21} cm$^{-2}$, and 
marginal evidence of dust extinction \mbox{$A_{V}=0.47\pm0.25$}. 
The simple power-law fit provides a good description of the dataset (W-stat=355 for 371 d.o.f.),
suggesting that optical and X-ray emission belong to the same
spectral segment \mbox{($\nu_m<\nu_{opt}<\nu_X<\nu_c$)} of the synchrotron spectrum.
The lack of significant spectral variation across the temporal break $t_j$ is
consistent with the properties of a jet-break, and exclude alternative interpretations (e.g. cooling frequency).

By extrapolating the observed spectrum to radio energies, the predicted flux at $t$=$t_1$ is $\approx$10~mJy, 
two orders of magnitude higher than the radio measurement. This implies a spectral break between the optical and radio band,
and that the radio data belong to a different spectral segment ($\nu_{r}$$<$$\nu_m$).
By adopting the standard closure relations for GRB afterglows \citep{zhame04} we fixed the radio spectral index to $\beta_r=1/3$, and fitted the broadband SED with a smoothly broken power-law.  
We added to the model a systematic uncertainty in order to take into account the possible effects of interstellar scattering and scintillation at radio wavelengths. 
Although a proper estimate of the ISS fluctuations requires more complex modeling (see Section~\ref{sec:agmod}), 
at this stage we introduce an uncertainty of $\approx$30\%.
Our fit constrains the spectral peak to lie in the IR region at $\nu_{pk}$$\approx$9.3\ee{11}~Hz at 0.5~d.
At our second epoch, the radio measurements are only slightly lower than the extrapolation of the higher energy spectrum,
implying that the spectral peak moved close to the radio band. We estimate $\nu_{pk}$$\approx$37 GHz at 2.5~d, 
above the VLA frequencies. This shows that the observed radio, optical and X-ray emission remained in the same spectral regime,
thus the observed temporal break was not caused by spectral variations. 
Basic considerations on the spectral and temporal behavior of the afterglow
disfavor a wind-like environment, which would cause a steeper decay ($\alpha_{wind}$$\approx$1.5) of the pre-break X-ray afterglow. 
Our analysis also shows that the broadband spectrum evolved in time 
as $\nu_{pk}$\,$\propto$\,$t^{-2}$, and $f_{pk}$\,$\propto$\,$t^{-0.3}$.
As shown in Figure~\ref{fig:sed} (right panels), these decay rates are significantly steeper than the ones predicted by the spherical fireball model for a uniform medium, and 
are instead consistent with the spectral evolution of a collimated outflow. In particular, the slow decay of the peak flux strongly favors a narrow jet model
seen slightly off-axis. 

\newpage
\subsection{Host Galaxy Properties}\label{host}

GRB~140903A is located on top of a compact and red galaxy,  suggestive of an old system.
Based on the galaxy sky densities in the $r$-band \citep{yasuda01}, we estimated a small probability of a chance association, $P_{ch}$$\approx$3\ee{-4} \citep{bloom02,troja08},
and we therefore consider this galaxy as the GRB host. 
From our $r$-band measurement we derive a rest-frame absolute $B$-band magnitude $M_B$$\approx$-20.9~mag,
or $L_B \approx 0.8 L_{*}$ when compared to the luminosity function of galaxies at a similar redshift 0.2$<$$z$$<$0.4 \citep{willmer06}.
In order to characterize the galaxy's physical properties
we used the late-time ($t>$3~d) optical and IR data to build the host galaxy SED,
and ran a photometric fit with a grid of spectral templates within LEPHARE v.~2.2 \citep{ilbert06}. 
The templates were created using the stellar population synthesis libraries of \citet{bc03} 
with the Padova 1994 evolutionary tracks, and assuming the initial mass function from \citet{chab03}.
We adopted an exponential star formation history  with different $e$-folding times $\tau$, and included the contribution of emission lines following \citet{k98}.

\begin{figure}[!t]
\centering
\vspace{0.3cm}
\includegraphics[angle=0,scale=0.5]{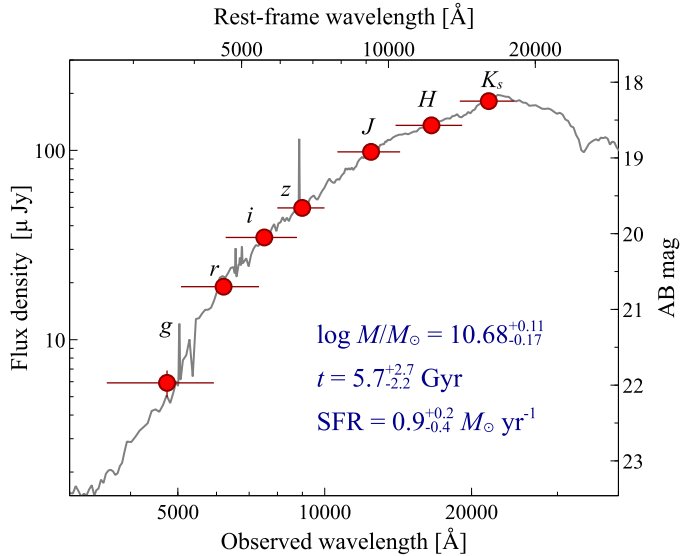}
\caption{Photometry of the galaxy hosting GRB 140903A. Data (filled circles) are corrected for Galactic extinction in the direction of the GRB.
The best fit stellar population synthesis model (gray curve), and its parameters are reported.}
\vspace{0.9cm}
\label{hostsed}
\end{figure}

Our results are shown in Figure~\ref{hostsed}. Our best fit model (gray curve) well reproduces the optical and NIR continua.
The best fit parameters for the galaxy template are: 
an intrinsic extinction $E_{B-V}$=0.25, 
solar metallicity, 
$e$-folding time $\tau$=500 Myr,  
stellar mass log(M/\msun )=10.61$\pm$0.15, 
an old stellar age $t$=4.1$^{+3.9}_{-2.3}$ Gyr, 
and a moderate star formation rate \mbox{SFR =1.0$\pm$0.3 \msun\,yr$^{-1}$} in agreement with the presence of
nebular emission lines in our spectra.

By using the extinction corrected H$\alpha$ line flux we infer a comparable value of SFR = 0.38$\pm$0.04 \msun\,yr$^{-1}$ \citep{k98} for a Chabrier IMF.
and a specific SFR of 0.47 $\pm$0.05 ($L/L_*$)  \msun\,yr$^{-1}$.  Based on the diagnostic  $F$([\ion{O}{3}] $\lambda$5007) / $F(H\beta)$$\sim$0.48 \citep{nagao06},
we estimate a super-solar metallicity 12 + log (O/H) $\approx$9.0$\pm$0.2, not unprecedented among short GRB host galaxies \citep{perley12}.


\section{Results}
\subsection{Origin of the X-ray emission}

The early X-ray afterglow of GRB~140903A is characterized by a period of fairly constant emission
lasting $\approx$4~hr. The shallow decay slope $\alpha_1$$\sim$0.2 is not consistent with 
a standard forward shock origin, and this is often considered a sign of 
prolonged energy injection into the blastwave \citep{zhang06, fan06, ctn11}. Indeed, it has been suggested that in a significant
fraction of GRBs the X-ray plateaus are originated by the internal dissipation 
of the engine driven wind rather than by shocks at an external radius. 
In this scenario, known as ``internal plateau"  \citep{troja07}, 
the forward shock component is sub-dominant, and the observed X-ray emission 
is {\it directly} powered by the central engine.
One of the most popular models invokes a newborn magnetar as the power source 
of the GRB and its afterglow: as the magnetar spins down,
it injects energy into the jet causing a period of nearly flat emission (the plateau),
followed by a steeper temporal decline with slope $\alpha \gtrsim$2 \citep{zhang01}.  
This rapid decay may mimic the presence of a jet-break, complicating the interpretation of the observed X-ray emission.

In the case of GRB~140903A, the standard expression for magnetic dipolar radiation \citep{shapiro83}
provides an excellent description of the X-ray dataset, as it can fully account for
the two salient features of the observed lightcurve -- a short-lived plateau and a final steep decay -- 
with the advantage of only two free parameters.  
However, due to the small radius at which the internal dissipation occurs, 
a bright optical and radio counterpart is not expected in these cases. 
Indeed a distinctive feature of `internal plateaus' is that they appear
as achromatic bumps visible in X-rays, but not at lower energies \citep{troja07,lyons10,rowlinson13}. 
Our SED analysis showed instead that X-ray, optical and radio data 
are consistent with being from the same emission component.  In particular, by considering that
the radio data lie above the self-absorption frequency $\nu_a$, we can derive a rough estimate of
the emitting radius $R \gtrsim$ 4\ee{16}~cm at $t$=0.5~d \citep{rodolfo13}, consistent with an external shock origin.
Moreover, the observed temporal and spectral indices ($\beta_{OX}$$\approx$0.7, $\alpha_2$$\approx$1.1) after the plateau phase
are in agreement with the canonical closure relations for $\nu_m<\nu_X<\nu_c$ and $p\approx$2.4.
Based on these considerations, we favor an external origin for the observed 
X-rays. In this scenario, the X-ray plateau is {\it indirectly} powered by the central engine
via sustained energy injection into the forward shock and, after the cessation of energy injection is communicated
to the shock front, the afterglow evolves in a standard fashion \citep{vaneerten14}. 
Therefore, the X-ray emission is not directly linked to the time history of the
central engine, instead it carries important information about the jet collimation, energetics, 
and  surrounding environment.

\subsection{Afterglow modeling}\label{sec:agmod}
We modeled the broadband dataset (from radio to X-rays) by using the standard prescriptions for an expanding spherical fireball, and the 
scaling relations for the post-jet-break evolution \citep{sari99}. We excluded from the fit the early time data ($t<t_{bk,1}$) as they
are affected by persistent energy injection. 
In our fit we implemented a routine to calculate the expected ISS modulation for each set of input afterglow parameters. 
By adopting the `NE2001' model \citep{cl02}, we derived a scattering measure SM~=~1.3\ee{-4}~kpc/m$^{-20/3}$ and 
a transition frequency $\nu_0$=8~GHz in the direction of GRB~140903A.  
Observations below this frequency could possibly be affected by strong scattering if the source size is smaller than the ISS angular scale, 
$\theta_{F0}$$\approx$1\,$\mu$as. At the GRB redshift this corresponds to an apparent fireball size $R_{\bot}$$\lesssim$2\ee{16}\,cm, 
which is likely the case at the early timescales here considered.
The derived ISS fluctuations were treated as a source of systematic uncertainty and added  in quadrature to the statistical errors when evaluating the fit statistics. 

We assumed a uniform circumburst medium with density $n_0$, and constant microphysical parameters $\epsilon_e$ and $\epsilon_B$. 
Under these assumptions, we did not find an acceptable fit to the data ($\chi^2$=65 for 43 dof),
mainly because the model predicts a much faster decay of the peak flux and peak frequency after the jet-break.
We attempted to model this effect by leaving the microphysical parameters free to vary in time as $\epsilon_e \propto t^{e}$
and $\epsilon_B \propto t^{b}$.
Although the fit formally improves for $b$$\approx$0.5, and $e$$\approx$0.2, it yields an unphysical solution $\epsilon_e>1$, 
and extreme values  for the blastwave kinetic energy and the jet opening angle. 
We considered this model an unrealistic description of the explosion,  and turned to a different interpretation to explain the observed properties. 

As shown in Figure~\ref{fig:sed} (right panels), the temporal evolution of the broadband spectrum appears roughly consistent
with a collimated fireball observed slightly off-axis. We therefore introduced in our model the effects of different viewing angles \citep{vaneerten10,vaneerten13}. 
This provides a better description of the observed data. The best fit parameters are 
an isotropic equivalent kinetic energy $E_{\rm K,iso}$=4.3$^{+1.2}_{-2.0}$\ee{52}\,erg, 
a circumburst density $n_0$=0.032$^{+0.14}_{-0.026}$\,\cm{-3}, 
and shock parameters $\epsilon_B$=2.1$^{+3.6}_{-1.4}$\ee{-4}, 
$\epsilon_e$=0.14$^{+0.19}_{-0.06}$. 
We derived a jet opening angle of $\theta_j$=0.090$\pm$0.012\,rad, 
and an observer's angle of $\theta_{obs}$$\approx$0.055\,rad.
These values are similar to the opening angles inferred from other candidate jet-breaks \citep{burrows06,coward12,fong15}.

\begin{figure*}[!t]
\centering
\vspace{0.cm}
\includegraphics[angle=0,scale=0.6]{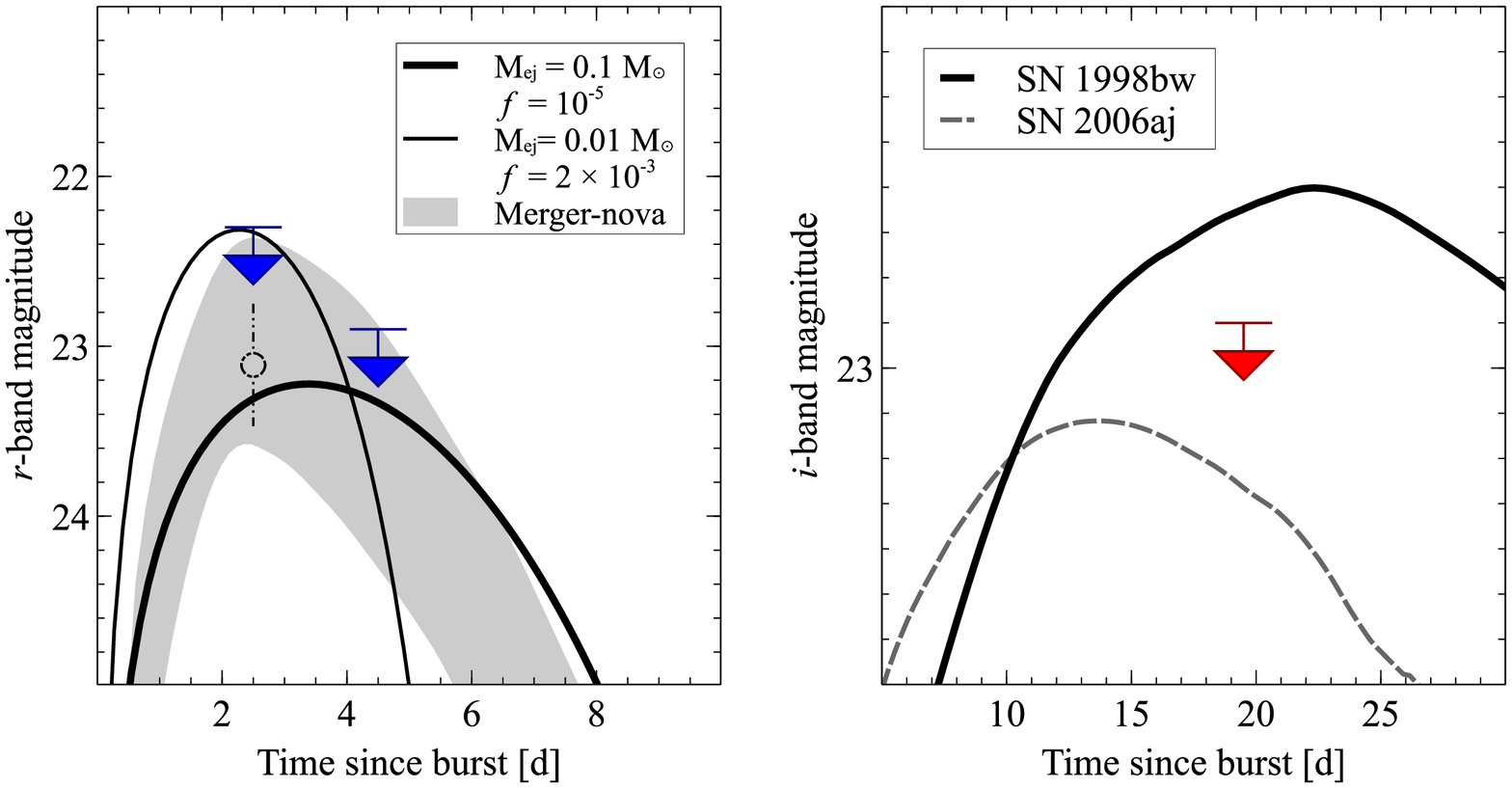}
\caption{ \textit{Left Panel:} Late-time $r$-band upper limits compared with theoretical light curves
of a macronova (solid lines) and a magnetar-driven merger-nova (shaded area). 
The dash-dotted symbol shows the low-significance signal visible in our DCT image at 2.5 days.
The macronova signal was derived by using the following parameters: 
a lanthanide-free opacity $\kappa$=1\,cm$^{2}$\,g$^{-1}$,  ejecta velocity $v$=0.1\,$c$,
ejecta mass $M_{ej}$=0.01\,\msun\ 
and a  rather high radioactive energy deposition $f$=2\ee{-3} (thin solid line); 
$\kappa$=1\,cm$^{2}$\,g$^{-1}$, $v$=0.1\,$c$,
$M_{ej}$=0.1\,\msun, 
and $f$=\e{-5} (thick solid line).
The merger-nova model was calculated by assuming a long-lived stable magnetar,  
and ejecta masses  \e{-4}\,\msun$<$$M_{ej}$$<$\e{-2}\,\msun\ (dashed line).
We applied to the models an extinction term as derived from the afterglow fit.
\textit{Right Panel:}  Late-time $i$-band observations compared with the extinction-corrected template light curves of  GRB-SNe: 
SN1998bw (solid line), and SN2006aj (dashed line).}
\vspace{0.1cm}
\label{minchionova}
\end{figure*}


\subsection{Constraints on SN-like transients}\label{sec:sn}

The possibility of an optical/IR transient rising a few days after the short GRB explosion is 
 the current focus of intense research \citep[e.g.][]{bk13,yu13,kasen15}. 
The detection and identification of such transients \citep[e.g.][]{tanvir13,yang15,jin15,jin16} would 
represent the smoking gun proof of short GRB progenitors, and
a powerful tool to search for electromagnetic counterparts of GW sources. 
We used our late-time observations to constrain some of the most promising models
as well as the presence of an emerging supernova.

As shown in Figure~\ref{minchionova} (left panel) our $r$-band upper limits  at 2.5~d and 4.5~d can constrain the
presence of a fast-rising and rapidly decaying transient, peaking in the optical a few days after the burst.
We considered two models: the classical \citet{lp98} macronova (or kilonova) powered by the radioactive decay of the ejecta (shaded area), 
and the more recent merger-nova \citep{yu13} powered by a long-lived magnetar (shaded area). 
Recent theoretical \citep{bk13} and observational \citep{tanvir13} results showed that the macronova emission is heavily suppressed at optical wavelengths
due to the high-opacity of the ejecta.  
Models for the late-time infrared emission \citep[e.g.][]{bk13}, although highly dependent on the input physics, generally predict a signal ($H \gtrsim$23~mag at $t$$\sim$4~d) well below the sensitivity of our observations. 
However, exceptions may occur if a small amount of lanthanide-free material is ejected
during the merger \citep{kasen15} or if the ejecta are re-energized by the central engine \citep{yu13}. 
The resulting transient spans a wide range of luminosities depending on the details of the explosion,
and our measurements can only constrain the bright end of the predicted values. 
For a \citet{lp98} macronova with a typical ejecta mass of $M_{ej}$=$0.01$\msun\ (thin solid line)
we can exclude only the extreme values of the $f$ parameter ($f$$>$2\ee{-3}), 
which measures the fraction of radioactive material converted into heat.  
Our limit is more interesting in the case of a larger ejecta mass of $M_{ej}$=0.1\msun,
for which we can exclude $f>$\e{-5} (thick solid line).
This is consistent with the most recent calculations of radioactive heating rate \citep{metz10,lippuner15}.

\citet{yu13} argued that, if the GRB central engine is a stable magnetar, the macronova luminosity could be boosted by several orders of magnitude. 
In this scenario, the main power source is the magnetar-driven wind rather than the radioactive decay energy. 
As shown in Figure~\ref{minchionova}, for a typical range of ejecta masses ($M_{ej}$$\lesssim$\e{-2}\,\msun)
the predicted signal of a merger-nova (dashed line) could be consistent with our observations. 

As mentioned in Section~\ref{dct}, we found marginal ($\lesssim$3\,$\sigma$) evidence of a signal in our observations 2.5 days post burst. 
The resulting magnitude, $r$=23.11$\pm$0.36, is above the predicted afterglow signal and, if real, would imply an optical rebrightening between our \textit{Gemini} observations at 1.5 d and the DCT observations at 2.5 d. When compared with the macronova predictions, this signal would require either an extreme value of the $f$-parameter 5\ee{-4}$<$$f$$<$\e{-3} for $M_{ej}$=0.01\,\msun, or a large ejecta mass, $M_{ej}$=0.1\,\msun, and $f\sim$\e{-5}, more typical of a NS-BH merger \citep{foucart14}.
The merger-nova predictions could instead reproduce the observed flux for ejecta masses $M_{ej}$\,$\approx$\e{-3}\,\msun, typical of NS-NS mergers \citep{baus13}.

Our last $i$-band observation, performed three weeks after the burst, is used to constrain the contribution of a possible SN. 
In Figure~\ref{minchionova} (right panel) we compare our upper limit with the light curves of SN1998bw and SN2006aj, associated to nearby long GRBs.
The templates were created by compiling data from literature \citep{galama98,ferrero06} and then corrected for cosmological effects and extinction in a standard fashion. 
Our limit ($M_V$$\gtrsim$-19~mag, rest-frame) is fainter than the emission expected from a SN1998bw-like explosion. 
Although our photometric dataset cannot exclude an event such as SN2006aj, we also note that the spectroscopic observations
do not show any evidence of broad absorption lines typical of GRB-SNe.

\section{Discussion}\label{sec:disc}

We have presented several lines of evidence linking GRB~140903A
to the class of short duration GRBs \citep{kouvel93}, and in support of the popular compact binary merger model.
Although characterized by a rather soft spectrum with photon index $\Gamma$$\sim$2, 
the GRB prompt emission displays a very short duration ($T_{90}$$\sim$0.3~s), 
negligible spectral lags, and a low luminosity ($L_{\gamma,\rm iso}$$\sim$\e{50}~erg~s$^{-1}$),
all key features of the class of short GRBs \citep{nb06, gehrels06}. 
The GRB afterglow was found on top of a relatively bright galaxy.  
Given the accurate afterglow localization, 
the probability of a chance alignment can be considered negligible ($P_{ch}$$\approx$0.03\%).
Moreover, the galaxy's properties (stellar mass, age, and metallicity) are broadly consistent 
with the population of short GRB host galaxies \citep{savaglio09,davanzo09,berger14}.
Both the environment and the lack of a bright SN (Section~\ref{sec:sn}) disfavor a massive star progenitor, 
and support instead the merger model for GRB~140903A. 

Direct evidence of a NS merger progenitor would be the detection of an r-process macronova \citep{lp98}. 
Our observations constrain only a limited range of the parameter space and, for the most likely values of ejecta masses and heating fraction, 
our upper limits are consistent with theoretical predictions. 
A  marginal detection in the residual image at 2.5 days could fit well the expected emission from a magnetar-driven macronova (or merger-nova; \citealt{yu13}). 
Unfortunately, given the low significance of the detection, the lack of confirmation in other bands, and the complexity of the field, 
we cannot exclude that the observed feature is an artifact of the subtraction process.
Although this does not allow us to draw any  robust conclusion on this particular event, it shows that rapid and deep observations of short GRBs
with large aperture telescopes are fundamental in order to pin down the possible onset of a macronova.

The most remarkable feature of this afterglow is the detection of an achromatic break at t$_j$$\approx$1~d
followed by a steep decay of the X-ray flux.  
Several mechanisms have been suggested to explain a rapid decay of the X-ray afterglow
\citep[e.g.][]{zhang01, troja07,vaneerten14}, although most of them predict a chromatic break preceding the steep flux decay. 
An achromatic break could be due to the cessation of energy injection. However, our analysis showed that
the pre-break afterglow is consistent with the standard closure relations \citep{zhame04} without energy injection.
We therefore interpret the observed break as evidence of a collimated outflow. 
Although early studies suggested the production of relatively wide outflows from NS mergers \citep{ruffert99,aloy05},
more recent works show that confinement from either the poloidal magnetic field \citep{rezzolla11} or
 the expanding cloud of ejecta \citep{naga14,duffell15} can produce a jet-like structure.
 Our observations of GRB~140903A add compelling evidence that, at least some, short GRBs are beamed into narrow jets.
  

In Section~\ref{sec:agmod}  we constrained the basic properties of the jet: an opening angle $\theta_j$$\approx$ 5\degree, 
an isotropic-equivalent energy release $E_{\rm K,iso}$$\approx$4\ee{52}\,erg, and a viewing angle $\theta_{obs}$$\approx$3\degree. 
Our modeling yields a blast-wave kinetic energy that is significantly higher than the observed prompt gamma-ray energy.
This would imply an unusually low radiative efficiency, $\eta_{\gamma} \approx$0.2\%.
However, since we observed the explosion slightly off-axis, the faint prompt emission could be due to a viewing angle effect:  
if the GRB jet is characterized by a compact central core and a steep radial gradient \citep{janka06},
an off-axis observer would indeed measure a dimmer and spectrally softer burst. 

The beaming factor $f_b$$\sim$250 has a direct impact on the GRB energy release and true event rate, and therefore  on the progenitor models. 
\citet{coward12} estimate the observed rate of short GRBs as $\sim$8\,Gpc$^{-3}$\,yr$^{-1}$. 
Collimation can boost this number up to $\sim$2\ee{3}\,Gpc$^{-3}$\,yr$^{-1}$, 
which is consistent with the conservative rate density of NS-NS mergers from \citet{abadie10}.  This would suggest that
most NS mergers successfully launch a short GRB, and that other systems, such as NS-BH or white dwarf binaries, 
do not contribute significantly to the observed GRB population. 
An important caveat to the above comparison between observations and progenitor models is that estimates of GRB jet angles are unavoidably biased 
by our observing strategy and limited sensitivity.  Narrowly collimated jets, if pointed toward us, are more likely to trigger \textit{Swift}
over a larger volume and to produce bright afterglows, allowing for the jet-break detection. On the other hand, 
wide outflows of comparable energy produce dimmer GRBs and afterglows, which are harder to detect and characterize. 
A proper assessment of the GRB event rate should properly account for these observational biases. 

The collimation-corrected energy release is $E$$\approx$2\ee{50}\,erg, 
which is in the typical range for short GRBs and lower than average long duration bursts \citep{cenko10,zhang15}.
Recently, \citealt{perna16} proposed a new mechanism to power a short GRB from a BH-BH collision. However, 
the low disc mass available in this system could only power a faint, low-luminosity transient, not consistent with the energetics measured in  our case. 
GRB~140903A was more likely produced by a merger event in which at least one of the two compact objects was a neutron star. 
According to the standard NS merger model, a stellar-mass black hole surrounded by a hot massive torus is formed after the merger. 
Energy  is extracted from this system through neutrino anti-neutrino annihilation or magnetically driven mechanisms.
Pair annihilation of neutrinos and antineutrinos can supply an energy deposition rate $L_{\nu \bar\nu}$$\lesssim$\e{51}\,erg\,s$^{-1}$
\citep{setiawan04,birkl07}, consistent with the energy budget of GRB~140903A.
Following the formalism of \citet{fan11}, we use the burst energetics to estimate a post-merger disc mass $M_{\rm disc}$$\approx$0.1\,\msun.
This is in agreement with numerical simulation of merging NS-NS and NS-BH binaries. 
If instead the outflow is driven by more efficient magnetic processes, the disc mass could be as low as \e{-3}\,\msun, suggesting
a  high-mass binary NS merger \citep{giacomazzo13a}.
An alternative scenario is the formation of a supra-massive and highly magnetized neutron star after the merger \citep{giacomazzo13b}. 
In this case, there are less robust predictions
connecting the central engine and the GRB observed properties. A general requirement is that the total energy release should not exceed
the maximum rotational energy of the newborn NS, $E_{rot}$$\approx$\e{53}\,($M_{NS}$/2\,$M_{\odot}$)$^{3/2}$ erg.  The burst energetics are well below this limit, 
and consistent with the proto-magnetar model.
A compact binary merger can therefore naturally explain the observed GRB properties, although the nature of the central engine 
and the energy extraction mechanisms remain uncertain.
Only future detections of gravitational wave radiation will be able to ultimately discriminate between these different scenarios.

\section{Conclusions}\label{sec:end}

We detected a temporal break in the X-ray afterglow light curve of the short GRB 140903A. The afterglow temporal decay was observed to steepen from $\alpha_1$$\sim$1.1 to $\alpha_2$$\sim$2.1,  suggesting the presence of a jet-break at $t_j$$\approx$1~d after the burst. Simultaneous observations at optical and radio wavelength showed that the break is achromatic. 
This disfavors a large set of models, including the  magnetar-powered `internal plateau', which are expected to produce a chromatic break. 
Instead we showed that the observed afterglow is consistent with the standard forward shock emission from a narrow jet expanding into a homogeneous medium.
We measure a jet opening angle of 5 deg, an observer's angle of 3 deg, and a total energy release of 2\ee{50}~erg. 
Several lines of evidences link this event to the popular NS merger scenario: the prompt gamma-ray emission, the environment, the lack of a bright SN, 
the energetics and rate of events. 
Our results show that NS mergers can produce highly collimated outflows.\\

\acknowledgements
The scientific results reported in this article are based in part on observations made by the Chandra X-ray Observatory.
Support for this work was provided by the National Aeronautics and Space Administration through Chandra Awards GO4-15072A and GO4-15067A issued by the Chandra X-ray Observatory Center, which is operated by the Smithsonian Astrophysical Observatory for and on behalf of the National Aeronautics Space Administration under contract NAS8-03060.
These results also made use of Lowell Observatory's Discovery Channel Telescope.
Lowell operates the DCT in partnership with Boston University, Northern Arizona 
University, the University of Maryland, and the University of Toledo. Partial 
support of the DCT was provided by Discovery Communications. LMI was built by 
Lowell Observatory using funds from the National Science Foundation (AST-1005313).
This paper is partly based on observations obtained at the Gemini Observatory, which is operated by the Association of Universities for Research in Astronomy, Inc., under a cooperative agreement with the NSF on behalf of the Gemini partnership: the National Science Foundation (United States), the National Research Council (Canada), CONICYT (Chile), Ministerio de Ciencia, Tecnolog\'{i}a e Innovaci\'{o}n Productiva (Argentina), and Minist\'{e}rio da Ci\^{e}ncia, Tecnologia e Inova\c{c}\~{a}o (Brazil).
Observations were also carried out with the 10.4 m Gran Telescopio Canarias installed in the Spanish Observatorio del Roque de los Muchachos of the Instituto de Astrofisica de Canarias in the island of La Palma (GTC59-14B) and with the 3.5m CAHA telescope at the German-Spanish Calar Alto Observatory operated by the IAA-CSIC. 
AJCT acknowledges support from the Spanish Ministry Projects AYA2012-39727-C03-01 and 2015-71718R.

\bibliographystyle{aa}
\bibliography{reference}

\end{document}